\documentclass[runningheads]{llncs}

\usepackage{eccv}

\usepackage{eccvabbrv}
\usepackage{multirow}
\usepackage{array} 
\usepackage{hhline} 
\usepackage{graphicx}
\usepackage{booktabs}
\usepackage[normalem]{ulem}
\useunder{\uline}{\ul}{}

\usepackage[accsupp]{axessibility}  

\usepackage[pagebackref,breaklinks,colorlinks,citecolor=eccvblue]{hyperref}

\usepackage{orcidlink}

\begin{document}

\title{Across-Game Engagement Modelling via Few-Shot Learning } 

\titlerunning{Across-Game Engagement Modelling via Few-Shot Learning}

\author{Kosmas Pinitas\inst{1}\orcidlink{0000-0003-0938-682X} \and
Konstantinos Makantasis\inst{2}\orcidlink{0000-0002-0889-2766} \and
Georgios N. Yannakakis\inst{1}\orcidlink{0000-0001-7793-1450}}

\authorrunning{Pinitas et al.}

\institute{Institute of Digital Games, University of Malta, Malta\and
Department of Artificial Intelligence, University of Malta, Malta \\
\email{\{kosmas.pinitas, konstantinos.makantasis, georgios.yannakakis\}@um.edu.mt}}

\maketitle

\begin{abstract}
Domain generalisation involves learning artificial intelligence (AI) models that can maintain high performance across diverse domains within a specific task. In video games, for instance, such AI models can supposedly learn to detect player actions across different games. Despite recent advancements in AI, domain generalisation for modelling the users' experience remains largely unexplored. While video games present unique challenges and opportunities for the analysis of user experience---due to their dynamic and rich contextual nature---modelling such experiences is limited by generally small datasets. As a result, conventional modelling methods often struggle to bridge the domain gap between users and games due to their reliance on large labelled training data and assumptions of common distributions of user experience. In this paper, we tackle this challenge by introducing a framework that decomposes the \emph{general} domain-agnostic modelling of user experience into several domain-\emph{specific} and game-dependent tasks that can be solved via few-shot learning. 
We test our framework on a variation of the publicly available \emph{GameVibe} corpus, designed specifically to test a model's ability to predict user engagement across different first-person shooter games. Our findings demonstrate the superior performance of few-shot learners over traditional modelling methods and thus showcase the potential of few-shot learning for robust experience modelling in video games and beyond.

\end{abstract}

\begin{figure*}[!tb]
\centering
\includegraphics[width=0.8\linewidth]{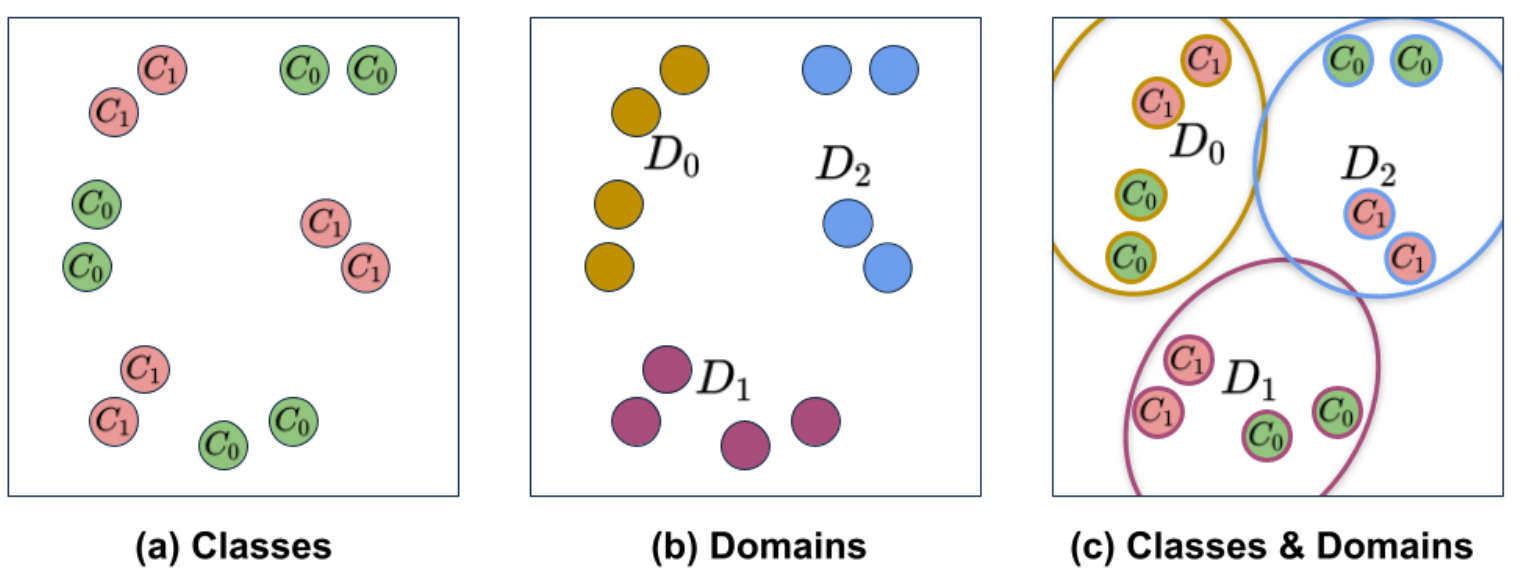}
\caption{Illustration of a classification problem---viewed as 2D plots of projected embeddings---containing three domains $D$ (yellow, magenta, blue) and two classes $C$ (green, pink). Such a problem is the prediction of user engagement classes (high vs. low) across several games (domains). Plot (a) showcases that the two classes cannot be easily separated across the entire dataset of domains. Plot (b) shows that points within the same domain are clustered together independently of their class. The rightmost plot (c) illustrates the method introduced in this paper combining both class and domain information: the method results in isolated domains allowing for classification within these more homogeneous groups.} 

\label{fig:concept}
\end{figure*} 

\section{Introduction}

Domain generalisation involves the creation of models that can perform well across various contextual factors (domains) within the same task (e.g., object recognition)\cite{zhou2022domain}. Building generalisable models becomes even more challenging when one has only limited samples per domain. 
Conventional methods may struggle to bridge the differences in data distribution between domains as they depend on large labelled training data and similar test data distributions. Various approaches have been developed over the years to address these challenges in computer vision subareas like object recognition \cite{li2017deeper}. Few-Shot Learning (FSL) \cite{wang2020generalizing} is a widely used method, enabling models to generalise to new data with minimal labelled samples by overcoming the inductive bias from the source domain distribution, thus adapting better to the target domain distribution.

Despite the rapid progress towards this direction, the area of domain generalisation within video games and user experience modelling remains underexplored. FSL is particularly well-suited for experience modelling because user experience annotations are inherently subjective, context-dependent and cannot be reliably collected in vast amounts \cite{yannakakis2023affective}. 
Moreover, video games, with their complex narratives, mechanics, levels and rich audiovisual stimuli, present unique challenges that FSL can address by using minimal yet diverse training samples to build robust and generalisable models \cite{yannakakis2018artificial}. Additionally, video games present a distinct opportunity for studying the dynamics of user experience, whether for players or viewers \cite{yannakakis2018artificial}. 
Unlike traditional media, video games feature numerous multimodal contextual factors varying widely across games---from visuals and audio, to text and navigation---that impact user experience directly. This creates, in turn, a vast and complex multidomain problem (as defined by different contextual factors, such as different games, game genres, and players) for any artificial intelligence (AI) method that wishes to generalise within. 

Motivated by the lack of few-shot domain generalisation studies in video games, in this paper, we propose a formulation for multidomain few-shot user experience modelling. Assuming consistent experience labelling behaviour, our framework treats each game as a separate domain, decomposing the multidomain problem into several non-overlapping domain-specific tasks (see Fig. \ref{fig:concept}) by incorporating domain-specific information into the labels. These tasks can then be solved simultaneously using few-shot learning approaches. To test our approach, we introduce the \emph{GameVibe} Few-Shot (GVFS) dataset, a few-shot variation of the \emph{GameVibe} \cite{barthet2024gamevibe} corpus. \emph{GameVibe} contains data from 30 First-Person Shooter (FPS) games annotated for viewer engagement, aimed at testing models' ability to generalise across contextual factors for the same task. Our initial study of FSL-based modeling focuses on \emph{viewer engagement}, which encompasses cognitive and affective user states like attention, interest, and enjoyment, offering a comprehensive understanding of user experience in games \cite{yannakakis2023affective}.

The paper presents several notable contributions. First, we propose an effective approach for turning multidomain classification tasks into few-shot learning tasks. Second, following the above-mentioned formulation, we devise GVFS, a few-shot learning variation of a publicly available game dataset specifically collected for learning models of viewer engagement that can generalise across various contextual factors. Third, we compare several few-shot learning approaches (e.g., metric learning and contrastive learning) against the conventional end-to-end engagement modelling approach, which is the standard practice in the field of affective computing and experience modelling. Finally, we employ various pretrained backbones and conduct experiments across 4 different scenarios, including 5- and 10-way, and 1- and 5-shot classification tasks. The results demonstrate that the proposed formulation is better suited for such problems since the few-shot learners achieve significantly higher accuracy values than the conventional domain-agnostic baseline in the vast majority of experiments performed. It should be noted that the proposed formulation is versatile and thus can be extended to any multidomain problem, beyond games, serving as a valuable tool for broader applications in machine learning and AI. 

\section{Background}

\textbf{Few-Shot Classification.} Few-shot learning is a paradigm focused on training models to classify new, unseen samples using only a few labelled examples. FSL methods are categorised into optimisation-based, metric learning, and hybrid approaches. Optimisation-based methods enable models to adapt quickly to new tasks with minimal data. Examples include MAML \cite{finn2017model}, which optimises model parameters for better adaptation; Reptile \cite{nichol2018reptile}, a MAML variant that reduces computational load by removing task-specific reinitialisation; and Meta-SGD \cite{li2017meta}, which optimises both parameters and learning rates. Metric learning approaches involve projecting inputs into a shared space where a metric is used to distinguish between classes. For instance, Prototypical Networks \cite{snell2017prototypical} learn a prototype for each class within a known metric space, ensuring that samples from the same class are closer to their prototype than to prototypes of other classes. Relation Networks \cite{sung2018learning} capture the relationship between data points by learning a deep distance metric that compares a small number of samples per iteration. Matching Networks \cite{vinyals2016matching} employ an attention mechanism to create a weighted nearest neighbour classifier using the support set of each learning iteration.  Recent work focuses at the intersection of FSL and contrastive learning (CL). Notably, Liu et al. \cite{liu2021learning}, used CL with noise contrastive estimation for few-shot image classification. ContrastNet \cite{chen2022contrastnet}, a CL framework, addresses representation and overfitting in text classification. Zhen et al. \cite{zheng2022few} proposed mixed-supervised hierarchical CL for aligning temporal clips. Jian et al. \cite{jian2022contrastive} combines supervised CL with masked language modelling in few-shot learning across various language tasks. In this work, we focus on domain generalisation within video games, an underexplored area in FSL research. We use FSL to model viewer engagement across multiple gaming contexts, and test our models' generalisation capacity by introducing the GVFS dataset. We treat each game as a distinct domain paving the way for future research in domain generalisation within this field.

\textbf{Modelling Experience in Games.} When it comes to games, experience modelling refers to the development of models that predict how a person behaves and feels while interacting with a game \cite{yannakakis2018artificial}. The majority of the early studies employed models focusing on hand-crafted features. Frommel et al. \cite{frommel2018towards} utilised the input from a graphics tablet and gameplay performance to predict the emotional state of players. Melhart et al. \cite{melhart2021towards} employed hand-crafted features that describe the gameplay context to learn general models of player arousal. Assuming that raw gameplay footage can be an effective elicitor to affect Makantasis et al. \cite{makantasis2019pixels} used CNN architectures to predict player arousal from gameplay footage. Lastly, Pinitas et al. \cite{pinitas2022rankneat} evolved the parameters of a preference learner to predict arousal in gameplay videos. Apart from the core dimensions of emotion, significant progress has been made in modelling more complex constructs of experience in games, such as engagement. Xue et al. \cite{xue2017dynamic} proposed a Dynamic Difficulty Adjustment framework to maximise player engagement. Huang et al. \cite{huang2019level} introduced a two-stage player engagement modelling method via Hidden Markov Models. Melhart et al. \cite{melhart2020moment} used chat logs as a proxy for engagement and predicted moment-to-moment gameplay engagement. Pinitas et al. \cite{pinitas2023predicting} employed pretrained CNN models and time-conditioning to predict long-term engagement in \emph{Tom Clancy’s The Division 2} (Ubisoft, 2020). Recently, Pan et al. \cite{10376289} proposed a CNN model for estimating streamers' engagement using gameplay footage, audio, and facial expressions. Unlike the above-mentioned studies, this paper explores the potential of FSL for robust experience modelling in video games with only few labelled samples. We evaluate several methods, including metric and CL ones, against end-to-end domain-agnostic engagement modelling methods. Our models use four different backbone architectures covering various FSL scenarios. The results demonstrate the superior performance of FSL against the conventional end-to-end approaches.

\section{Method}

\subsection{Problem Setting}

A primary contribution of this work is introducing a framework that improves learning from limited data with a high domain gap. Conventional end-to-end modelling methods struggle to generalise amid varying contextual factors within the same task, even in binary classification, especially when labels are noisy due to the subjective nature of engagement \cite{yannakakis2018ordinal}. Our framework exploits domain knowledge to decompose the classification problem into non-overlapping domain-specific sub-problems, with different classes falling under different domains. This ensures each domain has unique classes, which simplifies the learning process by reducing the variability and complexity within each sub-problem. Hence, Few-Shot Learning techniques can more effectively learn from the limited data available within each domain, as the model can focus on the specific features and patterns relevant to each unique domain-specific class.

\textbf{Modified Classification Objective:} \label{sec:modified_classification}
Let $f$ be a function that projects data from multiple domains $D_n$ with $n \in \{0...N\}$ into a common space of lower dimensions. We can decompose $f(\cdot)$ in the following manner:
\begin{equation}
f(x)=\sum_{n=1}^{N} f_n(x) \mathbf{1}_{D_n}(x)
\label{eq:domain_decomposition}
\end{equation}
\noindent
were $f_n(x)$ corresponds to the function within $n$-th domain $D_n$ and $\mathbf{1}_{D_n}(x)$ is the indicator function. It should be noted that $f_n(x)$ retains the core properties of $f(x)$ such as its invariance to specific transformations of $x$.  

Our objective is to learn a function $g(\cdot)$ that maps $f(\cdot)$ to the probability of discrete categories $y \in Y$, where $Y$ is the set of all possible classes. Consequently, the probability of class $y$ given $f(x)$ can be defined as follows:

\begin{equation}
g(f(x))=p(y|f(x))=   \sum_{n=1}^{N} p(y | f_n(x)) \mathbf{1}_{D_n}(x)
\label{eq:label_decomposition}
\end{equation}

\noindent
It is evident that the predicted probability distribution depends on the domain $D_n$. Thus we define $y_n$ to be the event $y | f_n(x)$ and consequently $p(y | f_n(x))=p(y_n)$ is the probability of class $y$ given the domain $D_n$. It is important to note that $y_n$ cannot occur outside of $D_n$ and due to that $p(y_n)=0$ for all $D_i\neq D_n$. As a result, the new classification objective is to learn a function $g(\cdot)$ that maps $f(\cdot)$ to discrete categories $y_n \in Y_{D_n}$, for $n=1,\cdots, N$, where $Y_{D_n}$ is the set of all possible $y_n$ classes within domain $D_n$.

In practice, we essentially need to define a set of distinct domain-specific classes with each class to exist only within its corresponding domain. To achieve this we define the relabelling function $R_Y$ as follows:
\begin{equation}
R_Y(y,n) = |Y|n+y
\label{eq:relabeling_function}
\end{equation}
\noindent
with $Y$ being the set of all possible labels of the domain-agnostic classification labels (e.g., $\{0,1\}$ for binary classification as illustrated in Fig. \ref{fig:concept}.a), $y\in Y$, $n$ a unique identifier of each domain and $|Y|$ is the cardinality of $Y$.  For a binary classification problem with 3 domains (Fig. \ref{fig:concept}.c), $|Y| = 2$ and $n \in \{0,1,2\}$. Consequently, the relabelled classes for these domains would be $\{0, 1\}$, $\{2, 3\}$, and $\{4, 5\}$ respectively. This ensures that the same class label in different domains is treated as a different label, thus creating non-overlapping domain-specific tasks.

It is worth noting that the domain identifier $n$ can be derived either by problem-specific knowledge (e.g., each game constitutes a different domain) or by applying a clustering algorithm on top of the initial projection function $f$.

\textbf{Learning From Limited Data:} An obvious drawback of the method described above is that it is not always possible to collect large amounts of data for every domain present within a dataset making the training of end-to-end classifiers infeasible. Additionally, even if we manage to train a conventional classifier, such a model will not be able to generalise in unseen categories of new domains. Few-shot classification allows models to learn and adapt to new, unseen categories using a few labelled examples. We define $D^{train}$, $D^{val}$, and $D^{test}$ as distinct and non-overlapping datasets for training, validation, and testing purposes. During each iteration, called an episode, data is sampled from $D^{train}$, $D^{val}$, or $D^{test}$. Each episode consists of $N$ classes (referred to as $N$-way) and $K$ samples per class (referred to as $K$-shot). Within an episode, a support set ($S$) and a query set ($Q$) containing labelled samples are defined from $D^{train}$, $D^{val}$ and $D^{test}$ during training, validation and testing, respectively. The model is trained to classify the samples in the query set using the information from the support set. The purpose of $D^{train}$, $D^{val}$, and $D^{test}$  (consisting of different domains) in FSL is to ensure the model can generalise well from a small number of examples, adapt to new tasks, and be evaluated fairly on its ability to handle unseen data with minimal supervision. Formally, the $i$-th sample in the support and query sets is represented, respectively, as $(x_i^s,y_i^s)$ and $(x_i^q,y_i^q)$. Note that $N$ (number of classes) and $K$ (number of samples per class) are hyperparameters that influence the difficulty of the few-shot classification setting. In particular higher values of $N$ increase the number of comparisons and lower values of $K$ reduce the amount of information available in the support set per episode. 

\subsection{Representation Components}
\label{sec:represemtation_components}

\begin{figure*}[!tb]
\centering
\includegraphics[width=0.7\linewidth]{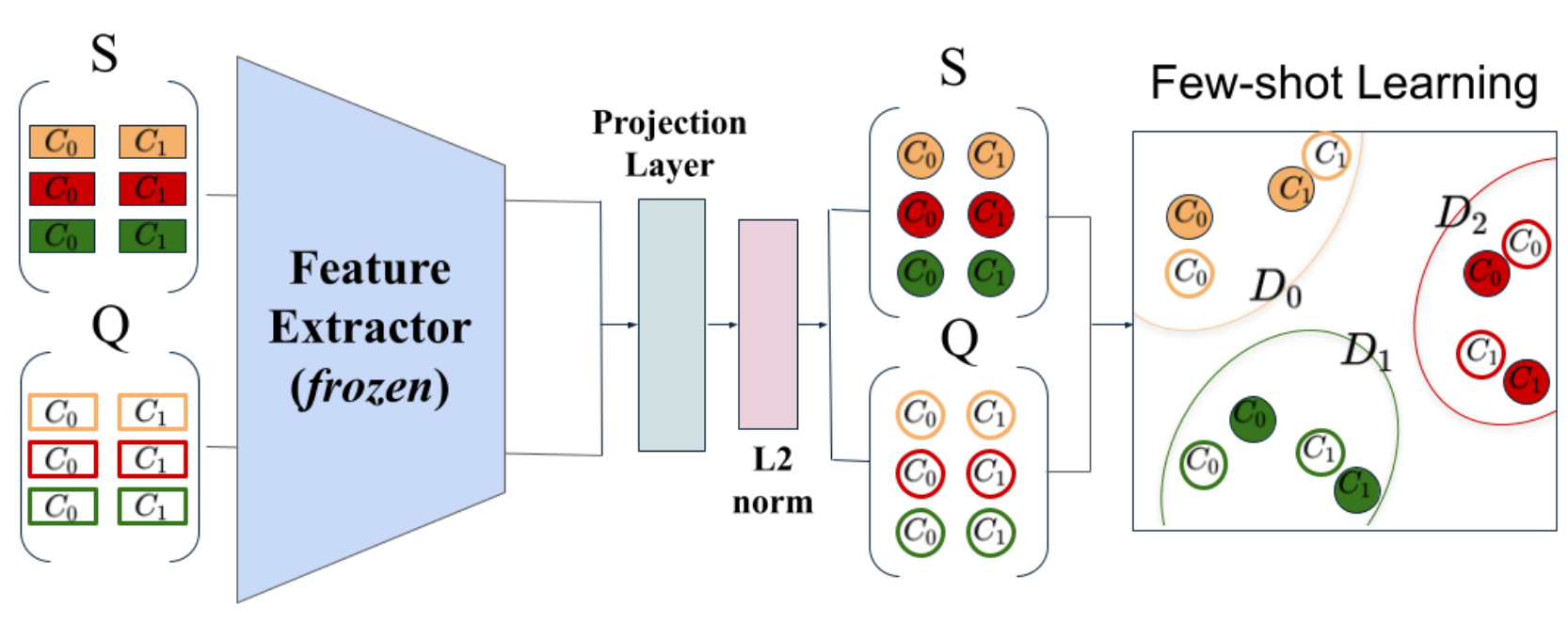}
\caption{A problem with 3 domains (yellow, red, green) and 2 classes ($C_0$ and $C_1$) per domain. $S$ and $Q$ represent the support and query sets, respectively. We first extract embeddings using a pre-trained frozen feature extractor. Following this step, we pass the extracted embeddings through a trainable projection layer and perform $L2$ normalisation. Finally, we optimise the few-shot losses using the resulting $S$ and $Q$. 
}
\label{fig:model_fig}
\end{figure*}

The overall methodology employed in this paper is illustrated in Fig. \ref{fig:model_fig}. In representation learning, an encoder is a neural network model that, after training, efficiently encodes input into low-dimensional, high-level representations. This work tests the capacity of learned representations to predict engagement from raw gameplay footage within FSL. We use four different backbones: one CNN-based and three Transformer-based, which are then finetuned for FSL tasks. The first encoder is I3D \cite{bertasius2021space}, using inflated 3D convolutions with an embedding size of 512. The Transformer encoders include the base versions of MVD \cite{wang2023masked}, VideoMAE \cite{tong2022videomae}, and VideoMAEv2 \cite{wang2023videomae}, each with a patch size of 16 and an embedding size of 768. All models are pretrained on Kinetics \cite{kay2017kinetics} and accept a tensor of 16 RGB frames as input. Additionally, a ReLU-activated layer, optimised by the FSL objectives described in Section \ref{method_objectives}, is added on top of the frozen backbones. This trainable layer has the same dimension as the output of the corresponding backbone, and $L2$-normalisation projects the embedding on the unit sphere.

\subsection{Learning Objectives} \label{method_objectives}
\label{sec:learning_objectives}

\textbf{Prototypical Network Loss:} The Prototypical Network (PN) loss \cite{snell2017prototypical} is being used to learn a metric space in which classification can be performed by computing the distances between the query samples and the prototypes derived for each of the $N$ classes within the support set. The PN loss is defined as

\begin{equation}
L_{PN} =  -\frac{1}{|Q|}\sum_{(x_i^q,y_i^q)\in Q}\sum_{n=1}^{N_q}  \mathbb{I}(y_i^q,n)\log(p_{\theta}(y_i^q=n|x_i^q)),
\label{eq:pn_loss}
\end{equation}
\noindent
where $p_{\theta}(y_i^q=n|x_i^q)=softmax(-d(f_\theta(x_i^q),c_n^s))$ is the probability of a query sample $x_i^q$ to fall into the class $n$, $c_n^s =\frac{1}{|S_n|}\sum_{(x_i^s,y_i^s)\in S} (y_i^s=n)f_\theta(x_i^s)$ is the prototype of class $n$, $d(\cdot)$ corresponds to the Euclidean distance, $f_\theta(\cdot)$ represents the learnable embedding functions, $|S_n|$ is the number of samples of class $n$ in the support set, and $|Q|$ is the cardinality of the query set. 

\textbf{Matching Network Loss:} Similarly to PN, the Matching Network (MN) Loss \cite{vinyals2016matching} aims to map an unlabelled example to a latent space defined by a small labelled set, enabling adaptation to new classes without fine-tuning. The optimisation objective uses a simple attention mechanism to weight sample distances between the support and query sets. The MN loss is defined as

\begin{equation}
L_{MN} = -\frac{1}{|Q|} \sum_{i=1}^{|Q|} \log p_{\theta}(y_i | x_i, S)
\label{eq:mn_loss}
\end{equation}
\noindent
where $|Q|$ is the number of query samples, $x_i$ is the $i$-th query sample, $y_i$ is the true label of the $i$-th query sample, and $S$ is the support set. Additionally  $p_{\theta}(y_i | x_i, S) = \sum_{(x_j,y_j) \in S} a(x_i, x_j) \cdot \mathbb{I}(y_j = y_i)$ where $a(x_i, x_j)$ is the attention mechanism defined as $a(x_i, x_j) = softmax(f_{\theta}(x_i) \cdot f_{\theta}(x_j))$ which is the probability of a query sample $x_i$ to be similar with the support sample $x_j$ when projected on the space $f_{\theta}(\cdot)$ defined by the support set $S$.

\textbf{Supervised Contrastive Loss:} The objective of supervised contrastive (SC) learning \cite{khosla2020supervised} is to derive representations that make samples with the same label (positive pairs) more similar and samples with different labels (negative pairs) more dissimilar. The minimisation of this loss yields distinct and separable representations for each class. Inspired by previous work in few-shot representation learning for image classification \cite{liu2021learning} we formulate SC as follows: 

\begin{equation}
L_{SC} = \frac{1}{|S|}\sum_{s\in S} \frac{-1}{|P_s^q|}\sum_{p \in P_s^q}\log\frac{\exp(r_s\cdot r_p/\tau)}{\sum_{q\in Q}\exp(r_s\cdot r_q/\tau)},
\label{eq:scl_loss}
\end{equation}
\noindent 
where $S$ is a set of all samples in the support set and $P_s^q$ is the set of only the query set samples that are assigned to the same class as $s$ while $q\in Q$ denotes any element in the query set. With $r_s$, $r_p$ and $r_q$ we denote the latent representations produced from a function $f_\theta$ that samples $x_s$, $x_p$ and $x_q$, respectively. $\tau$ is a non-negative temperature hyperparameter that transforms the representation similarity distribution.  Finally, $s$, $p$ and $q$ correspond to the index of the current support set sample, a query set sample positive to the current support sample and a sample in the query set, respectively. 

\section{Experiments}

\subsection{The GameVibe Few-Shot Dataset}

\begin{figure}[t]
\centering
\includegraphics[width=0.6\columnwidth]{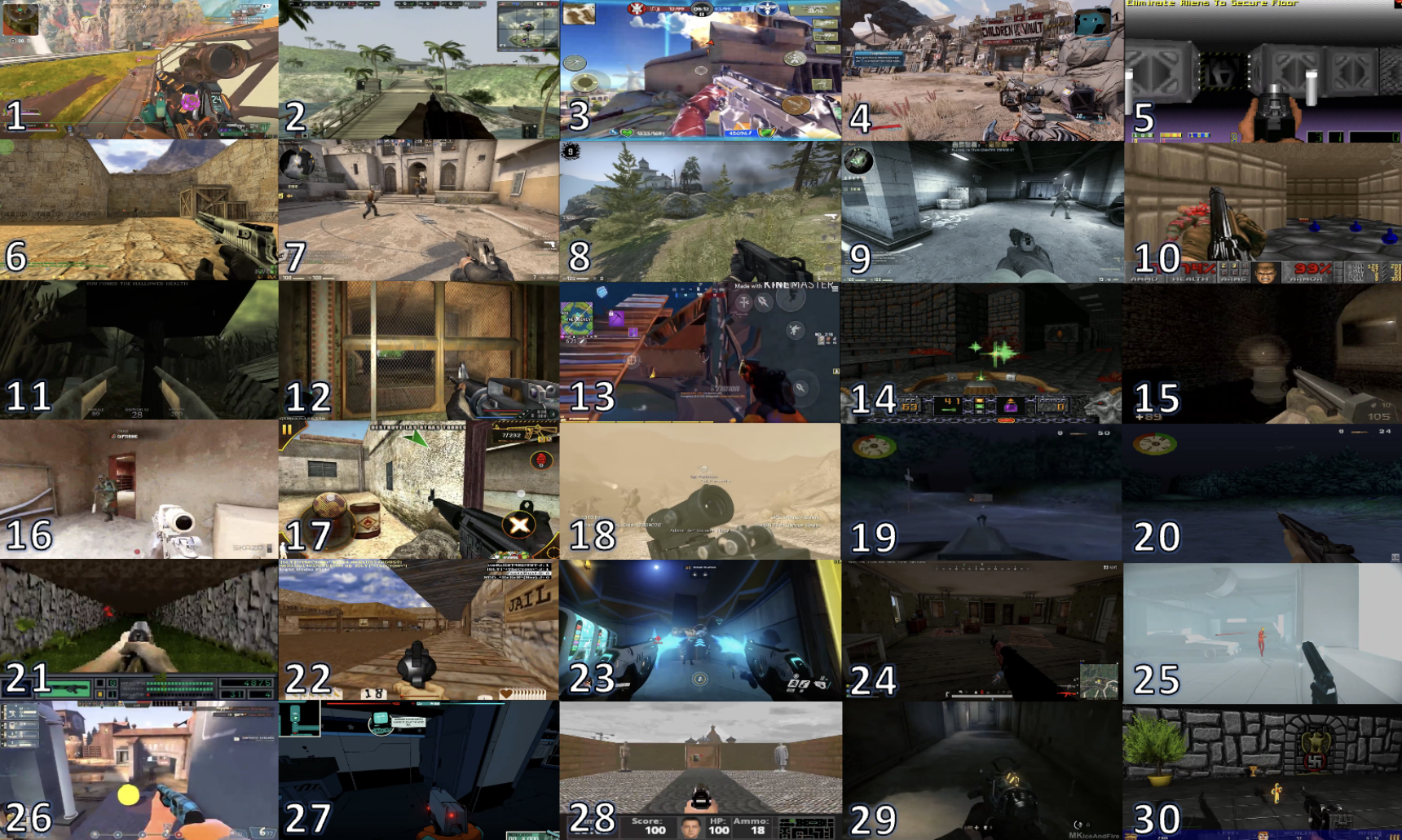}
\caption{Screenshots from the 30 different FPS games annotated for engagement. List of game titles: (1) Apex Legends; (2) Battlefield 1942; (3) Blitz Brigade; (4) Borderlands 3; (5) Corridor 7; (6) Counter-Strike 2016; (7) Counter-Strike 2018; (8) Counter-Strike 2019; (9) Counter-Strike: Global Offensive; (10) Doom; (11) Dusk; (12) Far Cry 1; (13) Fortnite; (14) Heretic; (15) Hrot; (16) Insurgency; (17) Modern Combat: Sandstorm; (18) Medal of Honor 2010; (19) Medal of Honor 1999; (20) Medal of Honor: Pacific Assault; (21) Operation Bodycount; (22) Outlaws; (23) Overwatch 2; (24) PUBG; (25) Superhot; (26) Team Fortress 2; (27) Void Bastards; (28) Wolfenstein 3D; (29) Wolfenstein New Order; (30) Wolfram Wolfenstein.}
\label{fig:games}
\end{figure}

The dataset is derived from the \emph{GameVibe} corpus \cite{barthet2024gamevibe}, a publicly available dataset of gameplay footage from 30 dissimilar commercial FPS games annotated for viewer engagement. The videos include audiovisual stimuli for engagement annotation, featuring diverse graphical styles and gameplay modes. Additionally, the videos contain only in-game sounds. All videos are limited to a maximum of 15 seconds of non-gameplay content, such as cut scenes or transition animations. Figure \ref{fig:games} illustrates the games of the corpus. The GameVibe corpus consists of four subcorpora (ids 1-4), each containing different gameplay clips from 30 FPS games and it is annotated by five randomly assigned participants, ensuring a variety of gameplay experiences and perspectives.

\textbf{Corpus Modalities:} The corpus includes two modalities: video frames and in-game audio. The video modality comprises high-resolution and low-resolution gameplay videos captured at 30Hz. Recent games have a resolution of $1280\times720$ pixels, while older games have $541\times650$ pixels. Each clip is 60 seconds long. The audio is stereo sound recorded at 44 kHz, extracted from the video. However, this study focuses solely on modelling engagement from raw gameplay footage. 

\textbf{Engagement Annotation:} Participants annotated 30 one-minute gameplay videos (one per game) based on the following engagement definition: \textit{``A high level of engagement is associated with a feeling of tension, excitement, and readiness. A low level of engagement is associated with boredom, low interest, and disassociation with the game''}. This definition of engagement is used due to its relevance and applicability within the specific context of FPS games \cite{pinitas2023predicting,barthet2024gamevibe}, where continuous attention and responsiveness are critical. The games were presented randomly to avoid habituation effects. Due to high cognitive load, annotating engagement for multiple games simultaneously is impractical \cite{metallinou2013annotation}. Short videos were chosen to balance annotation reliability and engagement stimuli richness. The 20 annotators (5 per subcorpus) were research staff and graduate students from the University of Malta. All annotation tasks were conducted in the same room with consistent conditions and equipment. Engagement annotations were collected with the RankTrace \cite{ranktrace} tool and were tested for reliability via annotator QA methods \cite{barthet2023knowing}. Data was collected and analysed respecting GDPR and national ethics guidelines \cite{melhart2023ethics} (see \textbf{ethical considerations} section below).

\textbf{Data Processing:} In this work, we focus on vision-based experience modelling in games following best practices from games research and affect modelling \cite{makantasis2019pixels,makantasis2022invariant,10376289}. Each session video is divided into non-overlapping 1-second time windows. To account for reaction time between gameplay and annotation, the input is shifted by 1 second relative to the annotation time window \cite{reaction_time}. Video segments are converted into RGB frames (30 per second). To reduce computational load, 16 RGB frames (224×224×3 pixels) are sampled at regular intervals within each window. Engagement traces (one per annotator, five per gameplay video) undergo min-max normalisation, scaling values to [0, 1] for each trace, and the median trace is derived to mitigate inter-annotator disagreement \cite{grewe2007emotions}. The resulting engagement trace is segmented into 1-second windows, and the average engagement value is calculated for each window.

\textbf{Defining Labels per Game:} Predicting user experience labels is challenging due to the subjective nature of such labels and the systematic reporting errors they might contain. Thus, following best practices from recent studies in the literature \cite{pinitas2022supervised,sanchez2022one,makantasis2023learning } we model engagement as categorical classification and we denote $e$ as the engagement value within a time window. Assuming that the annotators exhibit consistent behaviour across games, we initially define game-agnostic classification labels by binarising the $e$ values based on the median ground truth value of the entire set of affect annotation traces ($\bar{e}$) within a subcorpus (same participants within an annotation session). Consequently, the $i_{th}$ time window falls under the \emph{high engagement} and the \emph{low engagement} class, respectively, when ${e}_i > \bar{e} + \epsilon$ and ${e}_i < \bar{e} - \epsilon$. Notably, the threshold $\epsilon=0.1$ is used to eliminate ambiguous windows with annotation values close to $\bar{e}$ which may deteriorate the stability of the learning process (e.g., Fig. \ref{fig:concept}.a). Since each game constitutes a domain of its own, we apply the relabelling method discussed in Section \ref{sec:modified_classification} to construct game-specific labels of engagement. 

By utilising the unique game identifier $g_{ID} \in \{0,\ldots,29\}$ (domain identifier $n$ of Eq. \ref{eq:relabeling_function}), the set of (binary) labels derived in the previous step and the relabelling function of Eq. \ref{eq:relabeling_function} we define low and high engagement class labels for each game. For the game with $g_{ID}=0$ low and high engagement classes are represented by 0 and 1, respectively. For the game with $g_{ID}=1$ the same classes are represented by 2 and 3. In general, for a game with $g_{ID}=n$, low and high engagement classes are represented by $2n$ and $2n+1$, respectively. Since those classes are categories without a natural order, the game id value does not affect the output of our models. Finally, based on those labels we discard games that yield less than 10 samples per class since they don't allow for sampling $Q$ and $S$ sets from both classes and can lead to overinflated performance within the few-shot learning setting. The resulting dataset---\emph{GameVibe} Few-Shot---is the first dataset for few-shot experience modelling within video games. We refer to the four subcorpora of the dataset as GVFS\textsubscript{1}, GVFS\textsubscript{2}, GVFS\textsubscript{3}, and GVFS\textsubscript{4}, the key properties of which are summarised in Table \ref{dataset-stats-table}.

\begin{table}[!tb]
\centering
\caption{High-level statistics of the GVFS dataset. Each subcorpus includes 5 unique annotators. The train / valid / test columns refer to the number of games in the train validation and test set, respectively. Values within parentheses correspond to the number of distinct classes (2 per game).}
\label{dataset-stats-table}
\resizebox{0.7\textwidth}{!}{
\begin{tabular}{ccccc} 
\toprule
~ ~ subcorpus   & ~ ~ ~ ~\#samples & ~ ~ ~ ~\#games   & ~ ~\#train / valid / test & Binary Majority \\ 
\midrule
~ ~ GVFS\textsubscript{1}         & ~ ~ ~ ~ ~1054    & ~ ~ ~ ~ ~ ~23 & ~ ~ ~8 (16) / 8 (16) / 7 (14)    & 52.47\%       \\
~ ~ GVFS\textsubscript{2}         & ~ ~ ~ ~ ~1026   & ~ ~ ~ ~ ~ ~22 & ~ ~ ~8 (16) / 7 (14) / 7 (14) & 52.14\% \\

~ ~ GVFS\textsubscript{3}         & ~ ~ ~ ~ ~797    & ~ ~ ~ ~ ~ ~19 & ~ ~ ~7 (14) / 6 (12) / 6 (12) & 54.20\% \\

~ ~ GVFS\textsubscript{4}         & ~ ~ ~ ~ ~1186    & ~ ~ ~ ~ ~ ~27 & ~ ~ ~9 (18) / 9 (18) / 9 (18) & 50.34\% \\

\bottomrule
\end{tabular}}

\end{table}
\subsection{Experiment Protocol}

In this study, we compare models optimised by the losses mentioned in Section \ref{sec:learning_objectives}, using the $N$-way $K$-shot FSL evaluation setting. For each episode, we randomly sample $N$ classes and $K$ samples per class for both training and validation. Following best practices from earlier studies \cite{chen2022contrastnet,liu2021learning}, we report results on 5-way (5w) using both 1-shot (1s) and 5-shot (5s) settings, and we extend this evaluation protocol to the more challenging 10-way (10w) setting.

The models are trained using early-stopping after 10 epochs (40-200 episodes, depending on the experiment) without validation accuracy improvement, returning the best model based on validation accuracy. All models are optimised via SGD with a scheduler that halves the learning rate, $\alpha$, every 5 epochs (20-100 episodes). Preliminary experiments indicate optimal learning rate values within $\alpha \in (10^{-3},10^{-2})$. Hyperparameters are selected via greedy search on the validation set. Model performance is evaluated in terms of accuracy following the evaluation protocol of prototypical networks for fair comparison across FSL methodologies. We repeat the model training experiments 5 times and sample 200 test episodes per run. All reported significance tests are at a 95\% confidence interval (CI) with $p<0.05$. Finally, we test a conventional end-to-end baseline model that employs the same architecture as discussed in Section \ref{sec:represemtation_components}, specifically leveraging pretrained backbones with identical layers and hyperparameters. The end-to-end model is trained on binary game-agnostic labels by optimising the cross-entropy loss. This baseline predicts high and low engagement on unseen games, serving as the lower bound of performance.

\subsection{Results}

\begin{table*}[!tb]
\centering
\caption{\textbf{5-way few-shot experiments} (1-shot and 5-shot) across the GVFS subcorpora and on average. Mean accuracy on the End-to-end baseline (B), Matching Network Loss (MN) Prototypical Network Loss (PN) and Supervised Contrastive Loss (SC). Bold values indicate the highest accuracy obtained for each sub-corpus and backbone used. Underlined values denote methods whose accuracy is statistically equivalent to the highest accuracy obtained as determined by the 95\% CI.}
\resizebox{0.85\textwidth}{!}{
\begin{tabular}{cc|cccccccc|cc}
\hline
\multirow{2}{*}{\textbf{Backbone}} & \multirow{2}{*}{\textbf{Method}} & \multicolumn{2}{c}{{\ul \textbf{GVFS\textsubscript{1}}}} & \multicolumn{2}{c}{{\ul \textbf{GVFS\textsubscript{2}}}} & \multicolumn{2}{c}{{\ul \textbf{GVFS\textsubscript{3}}}} & \multicolumn{2}{c|}{{\ul \textbf{GVFS\textsubscript{4}}}} & \multicolumn{2}{c}{{\ul \textbf{Average}}} \\
 &  & \textbf{5s} & \textbf{1s} & \textbf{5s} & \textbf{1s} & \textbf{5s} & \textbf{1s} & \textbf{5s} & \textbf{1s} & \textbf{5s} & \textbf{1s} \\ \hline
\multirow{4}{*}{I3D} & B & 55.25 & 55.25 & 54.10 & 54.10 & 47.50 & 47.50 & 48.63 & 48.63 & 51.37 & 51.37 \\
 & MN & \textbf{91.21} & {\ul 84.71} & \textbf{91.34} & {\ul 83.30} & \textbf{89.02} & {\ul 79.87} & {\ul 92.88} & \textbf{85.42} & {\ul 91.11} & {\ul 83.33} \\
 & PN & \textbf{91.21} & {\ul 84.040} & \textbf{91.34} & {\ul 83.09} & {\ul 88.83} & {\ul 80.18} & \textbf{93.12} & {\ul 84.80} & \textbf{91.13} & {\ul 83.03} \\
 & SC & {\ul 91.19} & \textbf{84.80} & {\ul 89.75} & \textbf{83.76} & {\ul 89.02} & \textbf{82.53} & {\ul 91.83} & {\ul 85.02} & {\ul 90.45} & \textbf{84.03} \\ \hline
\multirow{4}{*}{MVD} & B & 49.27 & 49.27 & 54.26 & 54.26 & 52.30 & 52.30 & 46.64 & 46.64 & 50.62 & 50.62 \\
 & MN & \textbf{81.73} & {\ul 74.62} & \textbf{84.95} & {\ul 75.00} & {\ul 80.06} & \textbf{74.36} & \textbf{84.53} & {\ul 75.16} & \textbf{82.82} & {\ul 74.79} \\
 & PN & \textbf{81.73} & {\ul 74.13} & \textbf{84.95} & {\ul 74.69} & \textbf{80.07} & \textbf{74.36} & {\ul 84.52} & {\ul 74.31} & \textbf{82.82} & {\ul 74.37} \\
 & SC & 77.77 & \textbf{75.96} & {\ul 82.36} & \textbf{75.71} & {\ul 78.25} & {\ul 72.89} & 81.36 & \textbf{77.6} & 79.94 & \textbf{75.54} \\ \hline
\multirow{4}{*}{VideoMAE} & B & 53.68 & 53.68 & 49.45 & 49.45 & 52.92 & 52.92 & 49.25 & 49.25 & 51.33 & 51.33 \\
 & MN & {\ul 78.86} & 66.31 & 75.83 & 64.14 & 73.54 & {\ul 61.69} & 79.96 & 64.00 & 77.05 & 64.04 \\
 & PN & {\ul 77.95} & 64.80 & 75.83 & 62.96 & 73.35 & {\ul 61.51} & 79.96 & 63.64 & 76.77 & 63.23 \\
 & SC & \textbf{81.73} & \textbf{74.27} & \textbf{79.77} & \textbf{69.00} & \textbf{76.44} & \textbf{62.93} & \textbf{83.10} & \textbf{74.67} & \textbf{80.26} & \textbf{70.22} \\ \hline
\multirow{4}{*}{VideoMAEv2} & B & 50.06 & 50.06 & 54.73 & 54.73 & 50.45 & 50.45 & 45.93 & 45.93 & 50.29 & 50.29 \\
 & MN & {\ul 91.20} & \textbf{83.82} & {\ul 90.99} & {\ul 81.78} & {\ul 89.40} & {\ul 78.98} & {\ul 92.88} & {\ul 82.62} & {\ul 91.12} & {\ul 81.80} \\
 & PN & {\ul 91.89} & {\ul 83.2} & {\ul 90.93} & {\ul 82.11} & {\ul 89.40} & \textbf{80.27} & {\ul 92.95} & {\ul 83.73} & {\ul 91.29} & {\ul 82.33} \\
 & SC & \textbf{94.08} & {\ul 83.33} & \textbf{92.36} & \textbf{82.56} & \textbf{90.45} & {\ul 79.11} & \textbf{93.83} & \textbf{84.49} & \textbf{92.68} & \textbf{82.37} \\ \hline
\end{tabular}}

\label{tab:res_5_way}
\end{table*}

\begin{table*}[!tb]
\centering
\caption{\textbf{10-way few-shot experiments} (1-shot and 5-shot) across the GVFS subcorpora and on average. Mean accuracy on the End-to-end baseline (B), Matching Network Loss (MN) Prototypical Network Loss (PN) and Supervised Contrastive Loss (SC). Bold values indicate the highest accuracy obtained for each sub-corpus and backbone used. Underlined values denote methods whose accuracy is statistically equivalent to the highest accuracy obtained as determined by the 95\% CI.}
\resizebox{0.85\textwidth}{!}{
\begin{tabular}{cc|cccccccc|cc}
\hline
\multirow{2}{*}{\textbf{Backbone}} & \multirow{2}{*}{\textbf{Method}} & \multicolumn{2}{c}{{\ul \textbf{GVFS\textsubscript{1}}}} & \multicolumn{2}{c}{{\ul \textbf{GVFS\textsubscript{2}}}} & \multicolumn{2}{c}{{\ul \textbf{GVFS\textsubscript{3}}}} & \multicolumn{2}{c|}{{\ul \textbf{GVFS\textsubscript{4}}}} & \multicolumn{2}{c}{{\ul \textbf{Average}}} \\
 &  & \textbf{5s} & \textbf{1s} & \textbf{5s} & \textbf{1s} & \textbf{5s} & \textbf{1s} & \textbf{5s} & \textbf{1s} & \textbf{5s} & \textbf{1s} \\ \hline
\multirow{4}{*}{I3D} & B & 55.25 & 55.25 & 54.10 & 54.10 & 47.50 & 47.50 & 48.63 & 48.63 & 51.37 & 51.37 \\
 & MN & {\ul 85.18} & \textbf{78.22} & \textbf{84.53} & \textbf{77.53} & \textbf{82.65} & \textbf{75.47} & \textbf{87.16} & {\ul 78.29} & \textbf{84.88} & \textbf{77.38} \\
 & PN & \textbf{84.95} & {\ul 77.49} & {\ul 83.79} & {\ul 77.38} & \textbf{82.65} & {\ul 75.20} & \textbf{87.16} & {\ul 78.47} & {\ul 84.64} & {\ul 77.14} \\
 & SC & {\ul 83.46} & {\ul 77.02} & {\ul 82.69} & {\ul 77.25} & {\ul 82.61} & {\ul 75.11} & {\ul 86.08} & \textbf{78.56} & {\ul 83.71} & {\ul 76.99} \\ \hline
\multirow{4}{*}{MVD} & B & 49.27 & 49.27 & 54.26 & 54.26 & 52.30 & 52.30 & 46.64 & 46.64 & 50.62 & 50.62 \\
 & MN & \textbf{73.37} & {\ul 67.2} & \textbf{74.89} & {\ul 67.29} & \textbf{70.23} & {\ul 65.93} & \textbf{75.31} & {\ul 68.11} & \textbf{73.45} & {\ul 67.13} \\
 & PN & {\ul 73.13} & \textbf{68.09} & {\ul 74.88} & \textbf{67.53} & \textbf{70.23} & {\ul 65.93} & \textbf{75.31} & \textbf{68.18} & {\ul 73.39} & \textbf{67.43} \\
 & SC & {\ul 72.51} & {\ul 66.44} & {\ul 74.34} & {\ul 66.89} & {\ul 69.37} & \textbf{66.00} & {\ul 73.92} & {\ul 68.04} & {\ul 72.54} & {\ul 66.84} \\ \hline
\multirow{4}{*}{VideoMAE} & B & 53.68 & {\ul 53.68} & 49.45 & {\ul 49.45} & 52.92 & {\ul 52.92} & 49.25 & 49.25 & 51.33 & {\ul 51.33} \\
 & MN & {\ul 69.08} & {\ul 58.31} & 66.36 & {\ul 57.17} & 63.92 & {\ul 54.62} & 70.25 & {\ul 58.02} & 67.40 & {\ul 57.03} \\
 & PN & {\ul 68.36} & \textbf{59.73} & 66.36 & {\ul 56.95} & 63.92 & {\ul 54.93} & 70.25 & {\ul 57.11} & 67.22 & {\ul 57.18} \\
 & SC & \textbf{70.23} & {\ul 58.87} & \textbf{69.73} & \textbf{57.19} & \textbf{68.55} & \textbf{55.76} & \textbf{74.56} & \textbf{58.29} & \textbf{70.77} & \textbf{57.53} \\ \hline
\multirow{4}{*}{VideoMAEv2} & B & 50.06 & 50.06 & 54.73 & 54.73 & 50.45 & 50.45 & 45.93 & 45.93 & 50.29 & 50.29 \\
 & MN & {\ul 84.56} & {\ul 75.8} & {\ul 84.08} & {\ul 74.88} & 82.38 & {\ul 72.71} & {\ul 86.34} & {\ul 76.40} & {\ul 84.34} & {\ul 74.95} \\
 & PN & {\ul 84.27} & {\ul 76.67} & 83.72 & {\ul 74.88} & 82.16 & {\ul 72.13} & {\ul 86.33} & {\ul 76.22} & {\ul 84.12} & {\ul 74.98} \\
 & SC & \textbf{86.2} & \textbf{78.27} & \textbf{85.40} & \textbf{76.15} & \textbf{83.89} & \textbf{73.22} & \textbf{87.64} & \textbf{77.62} & \textbf{85.78} & \textbf{76.32} \\ \hline
\end{tabular}}

\label{tab:res_10_way}
\end{table*}

\noindent
The proposed methodology is tested extensively across the four different subcorpora of the GVFS dataset and four few-shot settings. Table \ref{tab:res_5_way} showcases the average accuracy of the models for the 5-way few-shot experimental setting (e.g., 5 classes are compared per episode) and the backbone architectures employed. It is evident that FSL significantly outperforms the end-to-end baseline (B) across both 5-shot and 1-shot experiments demonstrating the robustness of the proposed approach and the ability of the resulting FSL models to generalise across different domains within the same subcorpus. In particular, the SC method yields the highest accuracy in 20 out of 32 5-way experiments showcasing the reliability of the contrastive optimisation objective. However, all few-shot learners perform statistically on par in 22 out of 32 experiments indicating that the main source of performance improvement is the training framework (see Section \ref{sec:modified_classification}) and not the few-shot learning objectives themselves.  

Table \ref{tab:res_10_way} illustrates the average accuracy of the models for the 10-way few-shot experimental setting (e.g., 10 classes are compared per episode) and the backbone architectures employed. In the same vein, the few-shot learners significantly outperform the baseline models (B) across all 5-shot, and in the vast majority of the 1-shot experimental settings (e.g., 13 out of 16) further highlighting the capacity of the proposed approach to build models that are able to generalise well across domains within the same corpus. The few-shot learners and the end-to-end model performed statistically on par in the remaining 3 experiments, where VideoMAE was used as the backbone. Once again, the SC method marks the highest accuracy in 17 out of 32 10-way experiments. Moreover, all few-shot learners performed statistically on par in 27 out of 32 experiments further strengthening the efficacy of the proposed framework to generalise across domains with limited data regardless of the choice of the backbone and few-shot learning objective.

Importantly, the average performance of the models across the GVFS subcorpora showcases that accuracy improves significantly from 1-shot to 5-shot settings across all methods and backbones, reflecting the advantage of having more examples in few-shot learning. Furthermore, as the number of classes increases from 5-way to 10-way, a notable drop in performance is observed across all methods and backbones. This is due to the increased complexity and higher intra-class variability, making it harder for the models to distinguish between more classes with limited samples. Additional adversity comes from the subjective nature of experience annotation that inherently leads to noisy mappings between embeddings and labels. The baseline models (B) struggle to learn engagement patterns likely due to the significant domain gaps existent between different video games, thereby yielding poor generalisation. Unlike our approach, which uses domain-specific information to address these challenges, the baseline models assume a single distribution across games and thus fail to capture their unique features.

\section{Discussion}

This work introduced a framework for learning from limited data across multiple domains, focusing on engagement modelling in FPS games, where different games represent different domains. The framework decomposed the general problem (e.g., predicting engagement in unseen games) into domain-specific problems by incorporating domain-specific knowledge into engagement labels. This approach was tested on the \emph{GameVibe Few-Shot} dataset for few-shot experience modelling. We compared three FSL approaches with the conventional end-to-end method across the dataset's subcorpora. Results suggest that FSL models can effectively distinguish viewer engagement in unseen games, even with only one labelled sample per engagement level.

While the results demonstrate the method's efficacy in predicting viewer engagement in FPS games, this study does not consider individual player characteristics like personal interests, skill level or playing behaviour. Developing models that consider and adapt to individual player characteristics could yield a more personalised and accurate measure of engagement, thereby, enhancing the applicability and robustness of our approach across different gaming contexts. Future research should also attempt to extend this approach to other game genres, such as arcade, sport, puzzle or strategy games, which may exhibit more complex engagement patterns. The dissimilar FSL objectives we explored offered us valuable insights into the stability and generalisation of engagement models since the baseline model we compared them against follows the conventional classification training process in affective computing \cite{pinitas2022supervised, melhart2020moment, makantasis2019pixels}. Despite notable improvements in accuracy, there might be trade-offs in other aspects of model performance, such as increased training time and computational complexity, that we do not consider here. While focusing solely on accuracy as the primary evaluation metric is standard practice in the literature it limits our assessment's scope. Future work should provide a more detailed analysis of both the baseline and FSL performance while exploring additional metrics, such as robustness to noisy data and the impact of data quality. Addressing these aspects will further enhance the model's applicability and reliability in real-world gaming environments.

Additional directions for future research include the thorough testing of the framework's efficacy across other modalities, such as sound and graph-based structures. Another direction involves tackling more challenging problems by relaxing annotation consistency, requiring the model to handle visual diversity and label shifts. Extending the framework to address domain-incremental scenarios, where models detect novel domains and assign labels during training, is an obvious next step. We also plan to evaluate our method's effectiveness in generating representations for tasks like few-shot segmentation and preference learning. Despite these open directions, the framework remains versatile and applicable to any few-shot learning task in video games, including gameplaying agents and content generation \cite{yannakakis2018artificial}.

\textbf{Ethical Considerations:} The experiments were conducted on a publicly available dataset of gameplay footage and engagement annotations. Any personally identifiable information was replaced by untraceable ids. The Research Ethics Committee of the University of Malta approved the protocol. The dataset used contains no potentially offensive data, and the data processing has been described in detail to assist scientific reproducibility. Finally, to the best of our knowledge, our work does not contribute to the development of deceptive applications or the escalation of existing privacy or discriminatory issues. 

\section{Conclusion}

In this paper, we addressed the problem of few-shot domain generalisation in video games. We particularly introduced a new approach for multidomain few-shot experience modelling, utilising the \emph{GameVibe} Few-Shot dataset, a variant of the \emph{GameVibe} corpus, to test our models' generalisation capabilities across different contexts. We compared our method against existing modelling approaches and we thus tested several few-shot learning methods, such as metric learning and contrastive learning, and traditional end-to-end domain-agnostic approaches. Our experiments, across 4 few-shot classification settings and pretrained backbones, showed that few-shot learning methods outperform conventional techniques. These results highlight the potential of few-shot learning to effectively model noisy data that come from multiple domains. Thus our findings suggest that few-shot learning can advance AI and computer vision research within games and beyond, by reducing reliance on large annotated datasets, thereby making AI development more accessible and scalable even in highly subjective downstream tasks such as user engagement modelling.

\section*{Acknowledgements}
Makantasis was supported by Project ERICA (GA: REP-2023-36) financed by the Malta Council for Science \& Technology (MCST) 
through the FUSION: R\&I Research Excellence Programme. Pinitas and Yannakakis were supported by Project OPtiMaL funded by MCST through the SINO-MALTA Fund 2022.

\bibliographystyle{splncs04}
\bibliography{main}
\end{document}